\documentclass[useAMS,usegraphicx]{mn2e}
\usepackage{rotate}
\usepackage{rotating}
\usepackage{times}
\newif\ifAMStwofonts
\AMStwofontstrue
\def\gs{\mathrel{\hbox{\rlap{\hbox{\lower4pt\hbox{$\sim$}}}\hbox{$>$}}}}
\def\ls{\mathrel{\hbox{\rlap{\hbox{\lower4pt\hbox{$\sim$}}}\hbox{$<$}}}}

\def\chandra{{\it Chandra}}

\def\rosat{{\it ROSAT}}

\def\swift{{\it Swift}}
\def\xmm{{\it XMM-Newton}}

\def\et{{et al.\ }}

\def\mrk79{{Mrk~79}}
\def\pg08{{PG~0844+349}}

\def\3c{{3C~273}}
\def\rg{{\thinspace r_{\rm g}}}

\def\fvar{{F_{\rm var}}}
\def\chidof{{\chi^2_\nu/{\rm dof}}}

\def\delchi{{\Delta\chi^2}}
\def\feka{{Fe~K$\alpha$}}

\def\fekb{{Fe~K$\beta$}}

\def\oiii{{[O~\textsc{iii}]}}

\def\feii{{Fe~\textsc{ii}}}

\def\nh{{N_{\rm H}}}

\def\arcs{{\hbox{$^{\prime\prime}$}}}
\def\deg{^{\circ}}
\def\A{{\rm\thinspace \AA}}
\def\cm{{\rm\thinspace cm}}
\def\erg{{\rm\thinspace erg}}
\def\eV{{\rm\thinspace eV}}

\def\ga{{\rm\thinspace gauss}}

\def\keV{{\rm\thinspace keV}}
\def\km{{\rm\thinspace km}}

\def\Msun{\hbox{$\rm\thinspace M_{\odot}$}}

\def\s{{\rm\thinspace s}}
\def\ks{{\rm\thinspace ks}}
\def\ps{{\rm\thinspace s^{-1}}}

\def\cmps{\hbox{$\cm\s^{-1}\,$}}

\def\ergpscmps{\hbox{$\erg\cm^{-2}\s^{-1}\,$}}

\def\ergps{\hbox{$\erg\s^{-1}\,$}}

\def\kmps{\hbox{$\km\ps\,$}}

\def\pscm{\hbox{$\cm^{-2}\,$}}

\title[\pg08\ in an X-ray weak state]
      {
The quasar \pg08\ in an X-ray weak state
      }

\author[L. C. Gallo et al.]
       {L. C. Gallo,$^1$ 
	D. Grupe,$^2$
	N. Schartel,$^3$
	S. Komossa,$^4$
	G. Miniutti,$^5$  
	A. C. Fabian,$^6$ 
\newauthor	
and     M. Santos-Lleo$^3$
        \\ 
$^{1}$ Department of Astronomy and Physics, Saint Mary's University, 923 Robie Street, Halifax, NS, B3H 3C3, Canada \\
$^{2}$ Department of Astronomy and Astrophysics, Pennsylvania State
University, 525 Davey Lab, University Park, PA 16802, USA \\
$^{3}$ XMM-Newton Science Operations Centre, ESA, Villafranca del
Castillo, Apartado 78, E-28691 Villanueva de la Ca\~nada, Spain \\
$^{4}$  Max-Planck-Institut f\"ur extraterrestrische Physik,
Giessenbachstr., D-85748 Garching, Germany \\
$^{5}$ Centro de Astrobiologia (CSIC-INTA), Dep. de Astrofisica; LAEFF, PO Box 78, E-28691, Villanueva de la Ca\~nada, Madrid, Spain \\
$^{6}$ Institute of Astronomy, University of Cambridge, Madingley Road, Cambridge CB3 0HA\\
}
\date{Accepted. Received. }
\pagerange{\pageref{firstpage}--\pageref{lastpage}}
\pubyear{2010}
\begin{document}
\maketitle
\label{firstpage}

\begin{abstract}
In March 2009 the well-studied quasar, \pg08, was discovered with \swift\ to be 
in an  X-ray weak state.  A follow-up \xmm\ observation several weeks later generated a good quality
spectrum of the source, showing substantial curvature and spectral hardening.  In combination with
archival data at two previous epochs when the source was in a bright state, we examine the long-term spectral 
and timing properties of \pg08\ spanning nearly ten years and a factor of ten in brightness.
Partial covering and blurred reflection models are compared to the data at each flux state while attempting to 
maintain consistency between the various epochs.
In terms of the blurred reflection model, \pg08\ is in a reflection dominated state during the 2009 X-ray
weak observations, which can be understood in terms of light bending.  Moreover, the light bending scenario
can also account for the short-term (i.e. $\sim1000\s$) spectral variability in the source.
Other models cannot be decisively ruled out, but we note distinguishing features of the models that can be
explored for in higher signal-to-noise data from current and future observatories.

\end{abstract}


% --------------------------------------------------------------------------

\section{Introduction}
\label{sect:intro}

The X-ray spectra of active galactic nuclei (AGNs) can exhibit such significant differences from
one epoch to the next that it can become very difficult to extract meaningful constraints from single-epoch
spectra.  Multi-epoch data are particularly useful to constrain model components that
fluctuate on vastly different time scales, thereby 
removing excessive freedom in some fit parameters (e.g. Gallo \et 2010, 2007a,b; Grupe \et 2008; Zoghbi \et 2008)
and allowing for more insightful discussion.  

One attractive proposition is to catch AGN in a particularly low X-ray flux state (Gallo 2006).  
In principle, the low-flux state can provide the best perspective of the AGN environment, for example 
the nature of the accretion disc and associated relativistic features (e.g. Miller 2007) 
and/or the surrounding ionised plasmas (e.g. Longinotti \et 2008)
since the direct continuum (i.e. the power law component)
is substantially suppressed (e.g. Schartel \et 2010, Ballo \et 2008, Grupe \et 2008; Vignali \et 2008; Miniutti \et 2009).
Moreover, examining the spectral changes from high- to low-flux states can potentially reveal the 
origin of the variability, whether it be: light-bending (e.g. Miniutti \& Fabian 2004),
absorption (e.g. Turner \& Miller 2009), or beaming (e.g. Gallo \et 2010, 2007b).

Efforts to identify objects in the low X-ray flux state by monitoring AGNs with \swift\ and triggering a pointed
\xmm\ observation when an object is discovered to be in a low-state are proving advantageous 
(e.g. Grupe \et 2007, 2008).
We have recently applied this strategy to catch the type-I quasar, \pg08\ ($z=0.064$)
in an X-ray weak state.\footnote{An X-ray weak state implies that the AGN X-ray emission is significantly
diminished with respect to the AGN's UV flux.  This is not necessarily the same as a X-ray low flux state, which
makes no statement about the UV emission and typically refers to the brightness of the source relative to
previous X-ray observations. 
The X-ray state can be quantified by $\alpha_{ox}$ as we describe in Sect. 3.1.
In 2009, \pg08\ was in a X-ray weak and low-flux state.}

\pg08\ is a particularly interesting radio-quiet AGN.  Sufficiently bright in the optical to be
termed a quasar, it possesses strong \feii\ emission and weak \oiii\ like a narrow-line Seyfert~1 (NLS1),
though its FWHM(H$\beta$) ($2420\kmps$, Boroson \& Green 1992) is slightly above the oft-quoted upper limit  
for a NLS1 of $2000\kmps$ (Osterbrock \& Pogge 1985).  
Its black hole mass estimated from reverberation mapping is $9.24\pm3.81\times10^{8}\Msun$ (Peterson et al. 2004).
Several values can be estimated for its Eddington ratio based on the literature (e.g. Grupe \et 2010;
Vasudevan \& Fabian 2009), but the value is
apparently always sub-Eddington.

Early in the \xmm\ mission, an observation of \pg08\ garnered attention as it was found
in a historically high-flux state compared to previous X-ray observations (Brinkmann \et 2003).
Pounds \et (2003) reported absorption features in this same data set that they
interpreted as high-mass, relativistic outflows.  A subsequent reanalysis of the data 
along with a deeper observation by Brinkmann \et (2006) could not confirm the absorption features reported
by Pounds \et and attributed them to calibration uncertainties
(see also Vaughan \& Uttley 2008).  

In March 2009, \pg08\ was observed with \swift\ as part of an unrelated project (PI: Vanden Berk).
A quick look at the X-ray data revealed the quasar had entered a low-flux state.
A \swift\ Target of Opportunity observation (ToO) was initiated in April to determine if the AGN was still
X-ray dim.  As it was found to be so, an \xmm\ ToO was triggered a few days later
and it is these new data, in combination with the previous \xmm\
observations, that we report on here.

The observations and data reduction
are described in the following section.  In Section~3, the general properties of the data at each epoch are explored.
In Section~4 we fit the average spectrum at each epoch with various physical models to derive a
self-consistent interpretation for the long-term (yearly scale) changes.
We compare the model, consider the short-term (intra-observation) variability, and discuss our results
in Section~5.  We summarise our results in Section~6.

\section{Observations and data reduction}
\label{sect:data}

Snap-shot observations of \pg08\ were conducted with \swift\ (Gehrels \et 2004)
on two occasions in 2009, the details of which
can be found in Table~\ref{tab:obslog}.  The X-ray Telescope (XRT; Burrows \et 2005)
operated in photon counting mode
(Hill \et 2004) and the data were reduced with the task {\tt xrtpipeline v0.12.1}.
Source photons were extracted from a circular region of $47\arcs$ radius 
(corresponding to an encircled energy fraction of 90 per cent)
using {\tt xselect}.  The background events were selected from a nearby source-free
region with radius $236\arcs$.  Net source counts
 were low at both epochs (126 and 63) and
a spectrum was only created for the higher count spectrum (2009 March).
The UV-Optical Telescope (UVOT; Roming \et 2005) observed \pg08\ through
the various filters at both epochs.
The UVOT data were reduced and analyzed as
described in Poole \et (2008).
All magnitudes are corrected for Galactic reddening ($E_{\rm B-V}=0.037$; Schlegel \et 1998)
and are listed in Table~\ref{tab:uvlog}.

\pg08\ was observed with \xmm\ (Jansen \et 2001) in 2000 and 2009.
In 2001 the AGN was in the field-of-view during the observation of cluster of galaxies Vik~59. 
A summary of all X-ray 
observations is provided in Table~\ref{tab:obslog}.  
During the 2000 and 2009 observations the EPIC pn (Str\"uder \et 2001) was operated in
full-frame and large-window mode, respectively.  In 2001 the pn camera
did not collect data.  
The MOS (MOS1 and MOS2;
Turner \et 2001) cameras were operated in small-window mode in 2000 and 2009, and
full-frame mode in 2001.
The medium filter was in place for the first observation and the thin filter for the
other observations.
The Reflection Grating Spectrometers (RGS1 and RGS2; den Herder \et 2001)
also collected data during these observations, but will not be discussed here.
The data were of lower statistical quality when \pg08\ was in a low-flux state in 2009, and in 2001
the source was not detected as it was observed off-axis.  The 2000 RGS data were reported
by Pounds \et (2003).  
The Optical Monitor (OM; Mason \et 2001) operated in imaging mode and collected data in 
the $U$ filter in 2000
and various filters in 2009 (Table~\ref{tab:uvlog}).  No OM data
were collected in 2001 as the AGN was outside the OM field-of-view.
%---------------------------------------------------
\begin{table*}
\begin{center}
\caption{\pg08\ X-ray observation log.  
The observatory and instrument used to obtain the data is given in
column 2.  MOS refers to the combine MOS1 and MOS2.  The observation
ID is provided in column 3.  The UT start date and 
the good-time interval (after removal of background flaring intervals) 
are given in column 4 and 5, respectively. 
Column 6 provides the approximate
total number of source counts in the $0.5-10\keV$ band.  The MOS information in
columns 5 and 6 is for the 
combined MOS1 and MOS2.
}
\begin{tabular}{cccccc}                
\hline
(1) & (2) & (3) & (4) & (5) & (6) \\
Epoch   &  Telescope and  & Observation ID  & Start Date   &  Exposure & Counts\\
        & Instrument  &            & (year.mm.dd) &       (s) & \\
\hline
2000 & \xmm\ pn & 0103660201 & 2000.11.05 &  11940 & 23241 \\
     & \xmm\ MOS &           &            &  37930 & 48777 \\
2001 & \xmm\ MOS  & 0107860501 & 2001.10.08 & 127600 & 133484 \\
2009a & \swift\ XRT  &  & 2009.03.14 & 3618 & 126 \\
2009b & \swift\ XRT  &  & 2009.04.26 & 1614 & 63 \\
2009 & \xmm\ pn & 0554710101 & 2009.05.03 & 11700 & 3716  \\
     & \xmm\ MOS &           &            & 28870 & 2608  \\
\hline
\label{tab:obslog}
\end{tabular}
\end{center}
\end{table*}
%---------------------------------------------------
%---------------------------------------------------
\begin{table*}
\begin{center}
\caption{The optical/UV observations of \pg08. 
The observatory and instrument used to obtain the data is given in
column 2. The magnitude and associated statistical error is provide for each filter
in columns 3--8.  
While the OM and UVOT filter sets are not identical they are comparable (e.g. Grupe \et 2008).
The effective wavelength for each filter is given in $\A$ and shown in brackets
with the UVOT value preceding the OM.
All magnitudes are corrected for Galactic reddening of $E_{\rm B-V}=0.037$. 
}
\begin{tabular}{cccccccc}                
\hline
(1) & (2) & (3) & (4) & (5) & (6) & (7) & (8) \\
Epoch   &  Telescope & $V$  & $B$   & $U$ & $UVW1$ & $UVM2$ & $UVW2$\\
        &            & (5468/5235) & (4392/4050) & (3465/3275) & (2600/2675) & (2246/2205) & (1928/1984) \\
\hline
2000 & \xmm\ OM & &  & $13.92\pm0.01$ &  &  & \\
2009a & \swift\ UVOT  & $14.45\pm0.01$ & $14.79\pm0.01$ & $13.65\pm0.01$ & $13.66\pm0.01$ & $13.61\pm0.01$ & $13.64\pm0.01$ \\
2009b & \swift\ UVOT  & $14.49\pm0.02$ & $14.79\pm0.01$ & $13.68\pm0.01$ & $13.67\pm0.01$ & $13.55\pm0.02$ & $13.63\pm0.01$ \\ 
2009 & \xmm\ OM & $14.54\pm0.01$ & $14.98\pm0.01$  & $13.95\pm0.01$ & $13.71\pm0.01$ & $13.79\pm0.01$ & $13.81\pm0.02$ \\
\hline
\label{tab:uvlog}
\end{tabular}
\end{center}
\end{table*}
%---------------------------------------------------

The \xmm\ Observation Data Files (ODFs) from all observations
were processed to produce calibrated event lists using the \xmm\ 
Science Analysis System ({\tt SAS v10.0.0}).
Unwanted hot, dead, or flickering pixels were removed as were events due to
electronic noise.  Event energies were corrected for charge-transfer
inefficiencies.  
Light curves were extracted from these
event lists to search for periods of high background flaring.
Significant background flares were detected in 2000 and 2001, and those 
periods have been neglected.
The total amount of good exposure is listed in Table~\ref{tab:obslog}.
Source photons were extracted from an annulus around the source to mitigate the effects of
pile-up in the central region during 2000 and 2001.  Pile-up was negligible in
the 2009 observations and a circular extraction region that included all of the source was used.
The background photons were extracted from an off-source region on the same CCD.
Single and double events were selected for the pn detector, and
single-quadruple events were selected for the MOS.
EPIC response matrices were generated using the {\tt SAS}
tasks {\tt ARFGEN} and {\tt RMFGEN}.  
The MOS and pn data at each epoch were compared for consistency and determined to
be in agreement within known uncertainties (Guainazzi \et 2010).  
The MOS1 and MOS2 spectra were combined to produce a single MOS spectrum at each
epoch.

The \xmm\ X-ray spectra were grouped such that each bin had a signal-to-noise
ratio of at least 7 in 2000 and 2001, and 5 in 2009.  
Spectral fitting was performed using {\tt XSPEC v12.5.0}
(Arnaud 1996).
All parameters are reported in the rest frame of the source unless specified
otherwise.
The quoted errors on the model parameters correspond to a 90\% confidence
level for one interesting parameter (i.e. a $\Delta\chi^2$ = 2.7 criterion).
A value for the Galactic column density toward \pg08\ of
$3.39 \times 10^{20}\pscm$ (Elvis \et 1989) is adopted in all of the
spectral fits.
K-corrected luminosities are calculated using a
Hubble constant of $H_0$=$\rm 70\ km\ s^{-1}\ Mpc^{-1}$ and
a standard flat cosmology with $\Omega_{M}$ = 0.3 and $\Omega_\Lambda$ = 0.7.

% --------------------------------------------------------------------------

\section{A first-look at the data}

\subsection{The UV/X-ray spectral energy distribution}
\label{sect:sed}

The 2000 observation of \pg08\ represents a relatively typical state
for the AGN.  The optical-to-X-ray spectral slope
(i.e. $\alpha_{ox}$; Tananbaum \et 1979) was approximately $-1.4$ and was
within expected values given its $2500\A$ luminosity (Just \et 2007), 
where we have estimated the $2500\A$ values directly from the shape of the
spectral energy distribution (SED) (Grupe \et 2010).
In 2001 \pg08\ was slight brighter in the X-rays, but no OM data are available.
Assuming the same optical flux as in 2000, the 2001 slope would be $\alpha_{ox}\approx-1.3$
% --------------------------------------------------------------------------
\begin{figure}
\rotatebox{270}
{\scalebox{0.32}{\includegraphics{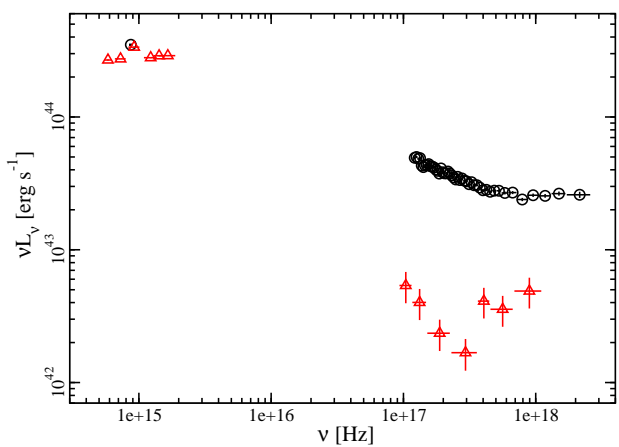}}}
\caption{The UV-to-X-ray SED of \pg08\ at two separate epochs.  In 2000 
(black circles) \xmm\ observed the AGN in a normal flux state.  In March 2009
(2009a, red triangles) \swift\ observed \pg08\ in an X-ray weak state.
The 2009a SED is representative of the spectral shape as it was observed again in
April and May 2009.
There was only one optical filter ($U$) used in 2000.  
}
\label{fig:sed}
\end{figure}
% --------------------------------------------------------------------------
In March 2009,
\swift\ detected a substantially lower X-ray flux, while no significant change
was measured in the optical.  Consequently the measured $\alpha_{ox}$
was steeper ($\alpha_{ox}\approx-1.86$) and the object was in an X-ray weak
state.  In Figure~\ref{fig:sed} the UV/X-ray SED
of \pg08\ in 2000 and March 2009 (i.e. 2009a) is shown.  The follow-up \swift\
observation in April 2009 revealed similar fluxes indicating a continued
X-ray low state so an \xmm\ Target of Opportunity (ToO) was triggered.
\xmm\ also caught \pg08\ in an X-ray weak state.  The measured UV-to-X-ray
slope during this observations was $\alpha_{ox}\approx-1.80$ and 
$\Delta\alpha_{ox}=-0.39$ (where $\Delta\alpha_{ox}$ is the difference between
the measured and expected $\alpha_{ox}$; Just \et 2007).
Following Grupe \et (2010), the Eddington ratio is $L/L_{Edd}\approx0.1$ during
both X-ray weak and normal states.

% --------------------------------------------------------------------------

\subsection{X-ray variability }
\label{sect:xvar}

% --------------------------------------------------------------------------
\begin{figure}
\rotatebox{270}
{\scalebox{0.32}{\includegraphics{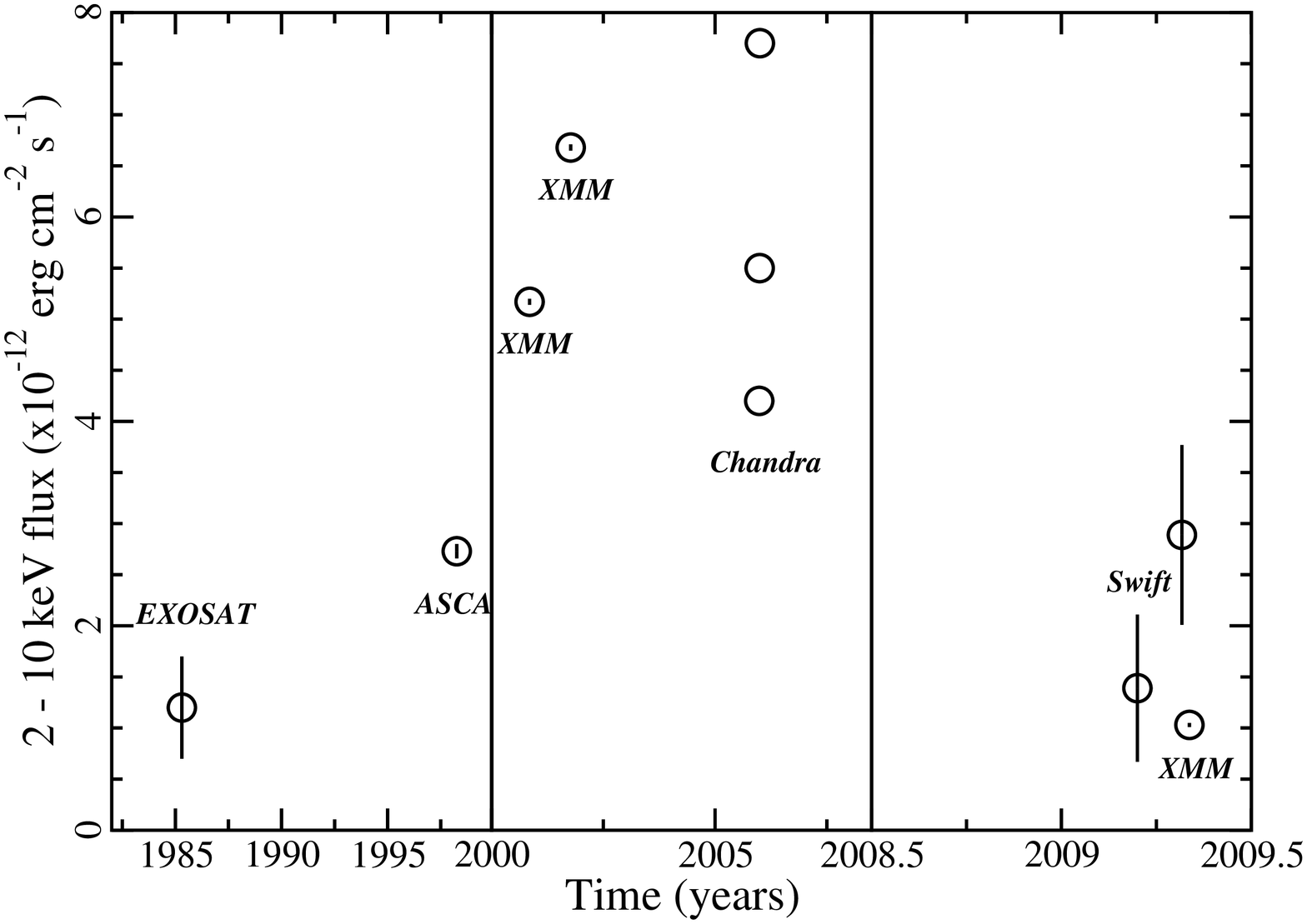}}}
\caption{The $2-10\keV$ light curve of \pg08\ spanning nearly 25 years. 
The instruments used to make the measurements are indicated beside the data points.
The \xmm\ and \swift\ fluxes come from this work and the \chandra\ values come from
Shu \et (2010).  The earlier flux values are taken from 
Wang \et (2000).  Departures into a low X-ray flux state are not uncommon for \pg08.
Error bars are included for all points except the for the \chandra\ data where none
were reported.  The time axis is continuous, but scales differently in each panel in order
to enhance clarity.
}
\label{fig:longlc}
\end{figure}
% --------------------------------------------------------------------------
Presented in Figure~\ref{fig:longlc} is the $2-10\keV$ flux of \pg08\ from all observations
over the last 25 years.  The light curve shows substantial variability and demonstrates that departures into
the X-ray low-flux state are common for \pg08.  
Not included in the figure are the 
\rosat\ data, which also show a factor of $6-10$ variability at low energies ($0.1-2.4\keV$) 
in observations separated by about 6-months
(Rachen \et 1996; Wang \et 2000).  Similar fluctuations are seen in the \xmm\ data
at energies between $0.5-2\keV$.

\begin{figure}
\rotatebox{270}
{\scalebox{0.32}{\includegraphics{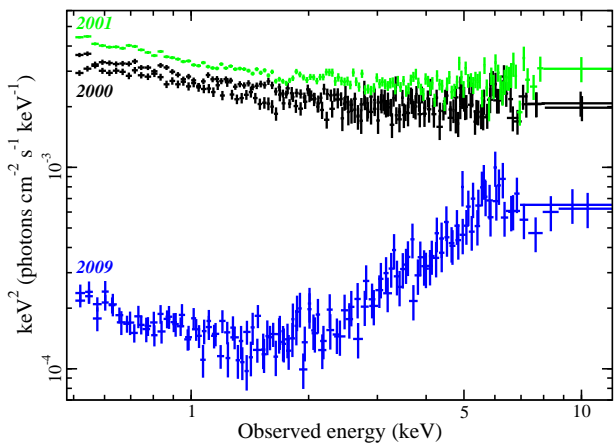}}}
\caption{The spectra from the three \xmm\ observations are shown after 
correcting for the effective area of the detectors.  The MOS and pn data are
shown with the same colour at each epoch to minimise confusion.  \pg08\ is in a low-flux
state in the 2009 and exhibits significant curvature in its spectrum (blue).
The different epochs are identified by colour: 2000 (black); 2001 (green); and
2009 (blue).
}
\label{fig:comp}
\end{figure}

\begin{figure}
\rotatebox{270}
{\scalebox{0.32}{\includegraphics{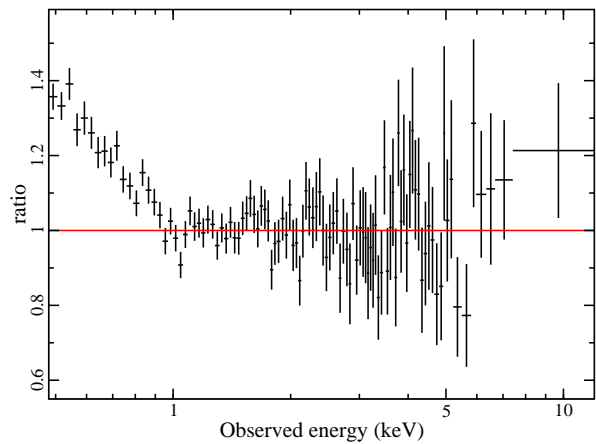}}}
\caption{The difference spectrum between the high-flux (2000) and low-flux (2009) data
is fit with a power law (corrected for Galactic absorption) between $1-10\keV$.  The
fit is then extrapolated over the $0.5-10\keV$ band.
}
\label{fig:diffs}
\end{figure}

The X-ray spectral variability from X-ray high- to weak- state are
depicted in Figure~\ref{fig:comp}.
% --------------------------------------------------------------------------
% --------------------------------------------------------------------------
The high-flux states show a rather smooth spectrum across the entire energy band.
In contrast the low-flux state exhibits substantial hardening and at 
lower energies \pg08\ is an order of magnitude dimmer than in the high-flux state. 
The difference spectrum between the high (2000) and low (2009) states is computed
revealing the shape of the variable component.  This spectrum is fit well with a power law
($\Gamma=2.4\pm0.03$) between $1-10\keV$, but deviates significantly when the fit
is extrapolated to lower energies (Figure~\ref{fig:diffs}).  The variable component
cannot be fit by a single power law across the entire band, thus indicating there is at least another
component contributing to the long-term variability.
% --------------------------------------------------------------------------
% --------------------------------------------------------------------------
In combination with the \rosat\ analysis by Wang \et (2000), the low-energy flux variations
may be generally more significant in \pg08\ than at high energies.

% --------------------------------------------------------------------------
\begin{figure}
\rotatebox{270}
{\scalebox{0.32}{\includegraphics{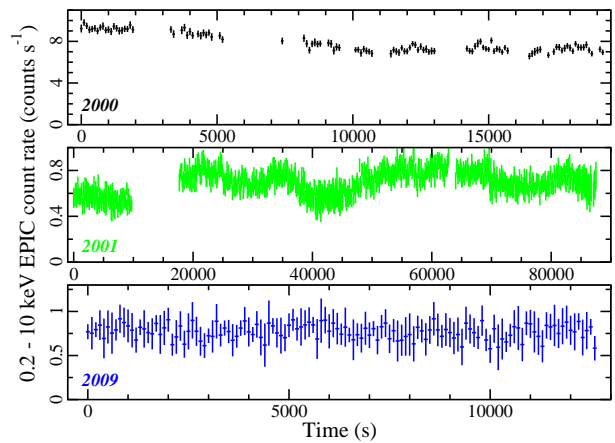}}}
\caption{The $0.2-10\keV$ EPIC light curves from the three observations of \pg08\
in $100\s$ bins.  
The counts from all EPIC instruments that collected data at each epoch have been added together.
N.B. Not all three cameras collected data at each observation so
the count rates on the vertical axes are not directly comparable between epochs
(see Table~\ref{tab:obslog}).
Zero on the time axes marks the start of each observation.  Note that the duration
of each observation is different.
}
\label{fig:lc}
\end{figure}
% --------------------------------------------------------------------------
The combined EPIC light curve from each \xmm\ observations is shown in Figure~\ref{fig:lc}.
Flux variations on rapid time scales (e.g. $\ls1000\s$) are present in the longer
and higher signal-to-noise light curves composing the two high state observations.
During the low-flux state the AGN is constant on the $\approx\pm15$~per cent level. 

Spectral variability is examined for in the longest observation (2001) and in the
low-flux state (2009) by calculating 
the fractional variability ($\fvar$) at various energies (e.g. Ponti \et 2004).  In 2001
\pg08\ was in a high-flux state corresponding to the brightest smooth spectrum
(green data) in Figure~\ref{fig:comp}.  The spectral variability is modest, but there is a trend
of increasing amplitude with energy.  The 2009 data are of lower quality as the variability is small
and the observation is short.  There 
is no detectable spectral variability within about $\pm15$~per cent.  The $\fvar$ spectra
are presented and discussed further in Section~\ref{sect:dis} and Figure~\ref{fig:fvar}.

Hardness ratio variability curves in various energy bands during the 2001 epoch 
show modest to low fluctuations with time.  There also appears to be no significant
correlation between hardness ratio fluctuations and count rate.  Similar conclusions were
reached by Brinkmann \et (2006).

\subsection{Phenomological X-ray spectral model}
\label{sect:xspec}
\begin{figure}
\rotatebox{270}
{\scalebox{0.32}{\includegraphics{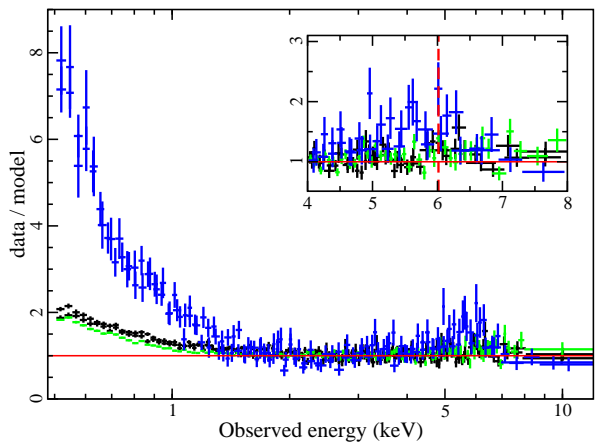}}
}
\caption{The residuals (data/model) remaining from fitting each spectrum
with a power law absorbed by Galactic absorption in the $2.2-4.5$ and $7-10\keV$ band, and then
extrapolating over $0.5-10\keV$.  An excess at low-energies ($<2\keV$) is prominent 
in all flux states (i.e. the soft excess). An excess at high-energies ($4-8\keV$) is more 
prominent in the low-flux state. The inset is an enlargement of the $4-8\keV$ region and
the axes are the same as on the large plot.  The vertical dashed line marks the rest-frame
$6.4\keV$.
Colours are as described in Figure~\ref{fig:comp}.
}
\label{fig:pofit}
\end{figure}

In this section we attempt to fit the spectra of \pg08\ with phenomological AGN models.
Fitting a power law modified by Galactic absorption to the $2.2-4.5\keV$
and $7-10\keV$ band, and then extrapolating over the $0.5-10\keV$ band generates the residuals
seen in Figure~\ref{fig:pofit}.  In the high-flux states the power law describes the data
well above $\sim2\keV$ and positive residuals are seen above the power law below $\sim2\keV$ (i.e. the 
soft excess).  
% --------------------------------------------------------------------------
% --------------------------------------------------------------------------
In 2009 the photon index of the power law fit is much flatter then in the high-flux state and
the soft excess appears much stronger.  There are also significant excess residuals
between $5-7\keV$ owing to the prominent spectral curvature in the low-flux state.

The soft excess is well fitted with a black body in addition to the underlying
power law.  The best-fit temperature is comparable in the high-flux states:
$110\pm5\eV$ in 2000 and $107\pm3\eV$ in 2001; but significantly higher in 2009
($133\pm6\eV$). All the temperatures fall in the typical range exhibited by unabsorbed
AGN (e.g. Gierli\'nski \& Done 2004; Crummy \et 2006).  The measured temperature is not 
correlated in an obvious manner with the black body luminosity (e.g. the highest temperature
corresponds to the intermediate luminosity) and
%, as in other NLS1 (e.g. Ponti \et 2010),  
a standard thermal accretion disc origin for the soft excess is unlikely.

A narrow ($\sigma=1\eV$) Gaussian profile is added to the fit to search for \feka\ emission
at $\sim6.4\keV$ arising from distant material (e.g. the torus).  
A marginal improvement over the power law model is seen at each epoch including such a feature, and
the measured energies and fluxes are consistent at all epochs.
Fitting a common feature at all epochs provides the most significant detection   
(Table~\ref{tab:gauss}).
When permitting the width of the line to vary, the fit prefers a broad feature likely
compensating for spectral curvature and/or emission due to He- and H-like iron.
While these high-ionisation lines may very well be present, they would not account
for the excess flux at $E<6.4\keV$.
Our measurements of energy and flux are consistent with the recent analysis of 
Shu \et (2010) who detected a significant $6.4\keV$ emission line in one-third of a
\chandra\ HETG observation of \pg08. 
%---------------------------------------------------
\begin{table}
\begin{center}
\caption{The measured energy and flux of an intrinsically narrow ($\sigma=1\eV$) Gaussian profile
that is added to the power law model in the $2.2-10\keV$ band.
Flux units are  \ergpscmps. The final row shows the improvement to the power law fit when
the two free parameters are added. 
}
\scalebox{0.8}{
\begin{tabular}{ccccc}                
\hline
         (1)&  (2)   & (3)  & (4)  & (5)      \\
Parameter   &  2000  & 2001 & 2009 & Combined \\
\hline
$E_{rest}$ (\keV) & $6.34\pm0.10$ & $6.44\pm0.06$ & $6.41^{+0.20}_{-0.10}$ & $6.41\pm0.06$ \\
%$\sigma$ (\eV) & $1^{f}$  &  & & \\
log$F$ & $-13.41^{+0.20}_{-0.37}$ & $-13.41^{+0.20}_{-0.38}$ &  $-13.65^{+0.18}_{-0.32}$ &  $-13.56^{+0.12}_{-0.18}$  \\
$\delchi$ & $7.7$ & $7.3$ &  $9.5$ &  $20.7$  \\
\hline
\label{tab:gauss}
\end{tabular}
}
\end{center}
\end{table}
%---------------------------------------------------

\section{Multi-epoch spectral analysis of \pg08}
\label{sect:multifit}

%---------------------------------------------------
\begin{table*}
%\begin{sidewaystable}[h]
%\begin{center}
\caption{The best-fit model parameters for the multi-epoch spectral modelling of \pg08.  
The model, model components and model parameters are listed in Columns 1, 2 and 3, respectively. 
Each subsequent column refers to a specific epoch.  
The inner ($R_{in}$) and outer ($R_{out}$) disc radius are given in units of gravitational radii 
($1\rg = GM/c^2$). 
The absorber covering fraction ($C_f$)
is given in percentage.
Values that are linked between epochs appear in only one column.  
The superscript $f$ identifies parameters that are fixed.
Fluxes are corrected for Galactic absorption and are in units of $\ergpscmps$.}
\centering
\scalebox{1.0}{
\begin{tabular}{cccccc}                
\hline
(1) & (2) & (3) & (4) & (5) & (6) \\
Model & Model Component &  Model Parameter  &  2000 & 2001 & 2009 \\
\hline
Blurred reflection & Power law & $\Gamma$ & $2.31\pm0.03$ & $2.24\pm0.02$ & $1.62^{+0.06}_{-0.02}$ \\
         &  &log $F_{0.5-10 keV}$ & $-11.084\pm0.001$ & $-10.933\pm0.001$ & $-12.420^{+0.061}_{-0.053}$ \\
\hline
 & Blurring  & $\alpha$    & $3.46^{+0.32}_{-0.41}$ & $4.04^{+0.36}_{-0.29}$ & $3.56\pm0.33$   \\
 &           & $R_{in}$ ($\rg$)   & $1.235^{f}$               &                        &               \\
 &           & $R_{out}$ ($\rg$)   & $300^{f}$                &                        &               \\
 &           & $i$ ($\deg$)  & $34\pm3$               &                        &                \\
\hline
 & Reflection  & $\xi$ ($\erg\cmps$)   & $119^{+21}_{-11}$  & $178^{+83}_{-25}$   & $50^{+15}_{-10}$       \\
 &             & $A_{Fe}$ (Fe/solar)   & $0.96\pm0.10$  &                     &    \\
% &             & $\Gamma$ & $2.31$ & $2.24$ & $1.62$ \\  
 &             &log $F_{0.5-10 keV}$ & $-11.343\pm0.022$ & $-11.054\pm0.013$   & $-11.850\pm0.026$ \\
\hline
 & Narrow Lines & $E_{FeK\alpha}$ (keV) & $6.40^{f}$ & & \\
 &              & $E_{FeK\beta}$ (keV) & $7.058^{f}$ & & \\
 &              & $\sigma$ (eV)     & $1^{f}$ & & \\
 &              & log$F_{FeK\alpha}$   & $-13.73^{+0.18}_{-0.31}$ & & \\
\hline
 &             Fit Quality & $\chidof$ & $1.05/474$ &   &  \\
\hline
\hline
Double Neutral & Power law & $\Gamma$ & $2.89\pm0.05$ & $2.96\pm0.04$ & $2.96\pm0.12$ \\
partial covering & &log $F_{0.5-10 keV}$ & $-10.423\pm0.002$ & $-10.230\pm0.002$ & $-10.714\pm0.001$ \\
\hline
 & Absorber 1& $\nh$ ($10^{22}\pscm$)   & $1.8^{+0.3}_{-0.2}$ & $2.1\pm0.2$ & $4.5^{+1.8}_{-1.3}$   \\
&           & $C_f$    & $0.45\pm0.04$ & $0.54\pm0.02$ & $0.80^{+0.05}_{-0.07}$   \\
% &           & $F_{abs}$  & $x.xx^{+0.xx}_{-0.xx}$ & $4.16^{+0.51}_{-0.40}$ & $4.83\pm0.11$   \\
\hline
 & Absorber 2& $\nh$ ($10^{22}\pscm$)   & $25^{+7}_{-5}$ & $26^{+6}_{-5}$ & $34^{+12}_{-7}$   \\
 &           & $C_f$    & $0.59\pm0.04$ & $0.58\pm0.04$ & $0.86^{+0.04}_{-0.05}$   \\
% &           & $F_{abs}$  & $x.xx^{+0.xx}_{-0.xx}$ & $4.16^{+0.51}_{-0.40}$ & $4.83\pm0.11$   \\
\hline
 & Narrow Lines & $E_{FeK\alpha}$ (keV) & $6.40^{f}$ & & \\
 &              & $E_{FeK\beta}$ (keV) & $7.058^{f}$ & & \\
 &              & $\sigma$ (eV)     & $1^{f}$ & & \\
 &              & log$F_{FeK\alpha}$   & $-13.60^{+0.20}_{-0.36}$ & & \\
\hline
 &             Fit Quality & $\chidof$ & $1.16/473$ &  & \\
\hline
\hline
Single Ionised & Power law & $\Gamma$ & $2.44\pm0.02$ & $2.67\pm0.02$ & $2.42\pm0.12$ \\
partial covering & &log $F_{0.5-10 keV}$ & $-10.619\pm0.003$ & $-10.528\pm0.002$ & $-11.037\pm0.009$ \\
\hline
 & Absorber 1& $\nh$ ($10^{22}\pscm$)    & $73\pm5$ & $13\pm2$ & $41\pm5$   \\
 &           & $C_f$    & $0.57^{+0.19}_{-0.09}$ & $0.57\pm0.02$ & $0.94\pm0.01$   \\
 &           & log$\xi$    & $2.2^{+0.2}_{-0.1}$ & $1.9^{+0.03}_{-0.13}$ & $1.95\pm0.03$   \\
% &           & $F_{abs}$  & $x.xx^{+0.xx}_{-0.xx}$ & $x.16^{+0.51}_{-0.40}$ & $x.83\pm0.11$   \\
\hline
 & Narrow Lines & $E_{FeK\alpha}$ (keV) & $6.40^{f}$ & & \\
 &              & $E_{FeK\beta}$ (keV) & $7.058^{f}$ & & \\
 &              & $\sigma$ (eV)     & $1^{f}$ & & \\
 &              & log$F_{FeK\alpha}$   & $-13.49^{+0.16}_{-0.25}$ & & \\
\hline
 &             Fit Quality & $\chidof$ & $1.36/476$ &   & \\
\hline
\hline
Double Ionised & Power law & $\Gamma$ & $2.60\pm0.07$ & $2.87\pm0.03$ & $2.70\pm0.10$ \\
partial covering & &log $F_{0.5-10 keV}$ & $-10.299\pm0.003$ & $-10.194\pm0.002$ & $-10.794\pm0.009$ \\
\hline
 & Absorber 1& $\nh$ ($10^{22}\pscm$)   & $168^{+8}_{-10}$ & $18\pm10$ & $18^{+5}_{-11}$ \\
 &           & $C_f$    & $0.75\pm0.10$ & $0.58^{+0.06}_{-0.04}$ & $0.91^{+0.03}_{-0.19}$   \\
 &           & log$\xi$    & $2.8\pm0.1$ & $0.2^{+0.5}_{-0.2}$ & $1.6\pm0.3$   \\
% &           & $F_{abs}$  & $x.xx^{+0.xx}_{-0.xx}$ & $x.16^{+0.51}_{-0.40}$ & $x.83\pm0.11$   \\
\hline
 & Absorber 2& $\nh$ ($10^{22}\pscm$)   & $5.8^{+1.2}_{-0.6}$ & $3.8\pm1.3$ & $39^{+28}_{-17}$   \\
 &           & $C_f$    & $0.39\pm0.09$ & $0.56\pm0.07$ & $0.64^{+0.11}_{-0.20}$   \\
% &           & log$\xi$    & $1.69\pm0.22$ & $1.92^{+0.18}_{-0.11}$ & $1.94\pm0.01$   \\
 &           & log$\xi$    & $0.9^{+0.6}_{-3.9}$ &  &    \\
% &           & $F_{abs}$  & $x.xx^{+0.xx}_{-0.xx}$ & $4.16^{+0.51}_{-0.40}$ & $4.83\pm0.11$   \\
\hline
 & Narrow Lines & $E_{FeK\alpha}$ (keV) & $6.40^{f}$ & & \\
 &              & $E_{FeK\beta}$ (keV) & $7.058^{f}$ & & \\
 &              & $\sigma$ (eV)     & $1^{f}$ & & \\
 &              & log$F_{FeK\alpha}$   & $-13.49^{+0.20}_{-0.39}$ & & \\
\hline
 &             Fit Quality & $\chidof$ & $1.06/469$ &  & \\
\hline
\label{tab:fits}
\end{tabular}
}
%\end{center}
%\end{sidewaystable}
\end{table*}
%---------------------------------------------------

The spectral changes from high- to low-flux state in \pg08\ are significant (e.g. Figure~\ref{fig:comp}).   In this
section we will attempt to model the entire data set with more physical models and in a self consistent
manner.  In shape, the low-flux spectrum of \pg08\ is reminiscent of other AGNs caught in a low-flux
state (e.g. 1H~0707--495 and Mrk~335).  The spectrum of such objects being modelled by 
blurred reflection or some form of partial covering.  All models include two narrow
Gaussian profiles with fixed energies and widths ($\sigma=1\eV$) that account for distant
\feka\ ($E=6.4\keV$) 
detected here and with \chandra\ (Shu \et 2010), and the undetected but expected \fekb\ ($7.058\keV$) emission line. 
The normalisation of the 
\fekb\ feature is fixed to 0.17 that of the \feka.  
All models and parameters
are shown in Table~\ref{tab:fits}.
Based on the statistically best-fit model (i.e. the ionised disc reflection) the $0.5-2\keV$ luminosities of \pg08\
(corrected for Galactic column density) are $\approx 5\times10^{42}$ and 
$10^{44}\ergps$ in the low- and high-flux state, respectively.  The corresponding $2-10\keV$
luminosities in the low and high state are $\approx10^{43}$ and $7\times10^{43}\ergps$, respectively.

\subsection{Ionised disc reflection }
\label{sect:refl}

Reflection from an ionised disc blurred for relativistic effects close to
the black hole is often adopted to describe the origin of the soft excess (Ross \& Fabian 2005; 
Ballantyne \et 2001), and has been successfully fitted to the spectra of unabsorbed AGN
(e.g. Fabian \et 2004; Crummy \et 2006).  In particular, the spectral complexity seen in the
X-ray weak state can be attributed to a reflection dominated spectrum.  That is when the direct power
law component seen by the observer is significantly diminished due to obscuration (e.g. Fabian \et 2002)
or light bending (Miniutti \&Fabian 2004). 

An intrinsic power law continuum plus blurred ({\tt kdblur})  reflection model is fit to the data of
\pg08.  The use of multi-epoch data permits linking of parameters that are not expected to vary over observable 
periods, for example: disc inclination ($i$), iron abundance ($A_{Fe}$),
outer disc radius ($R_{out}$).  To further simplify the fitting process the inner ($R_{in}$) and outer
disc radius are fixed at $1.235\rg$ and $300\rg$, respectively ($1\rg=GM/c^2$).  Allowing these parameters
to vary reproduced similar values, but did not significantly improve the fit quality.  The only blurring
parameter allowed to vary at each epoch is the emissivity profile of the disc, which is a
power law in radius ($J(r) \propto r^{-\alpha}$).  The 
ionisation parameter of the disc ($\xi=L_{x}/nr^{2}$, where $n$ is the hydrogen number density and $r$ is
the distance between the ionising source and the material), power law
continuum ($\Gamma$ and normalisation) and normalisation of the reflector are
permitted to vary at each epoch.
% --------------------------------------------------------------------------
\begin{figure}
\rotatebox{270}
{\scalebox{0.32}{\includegraphics{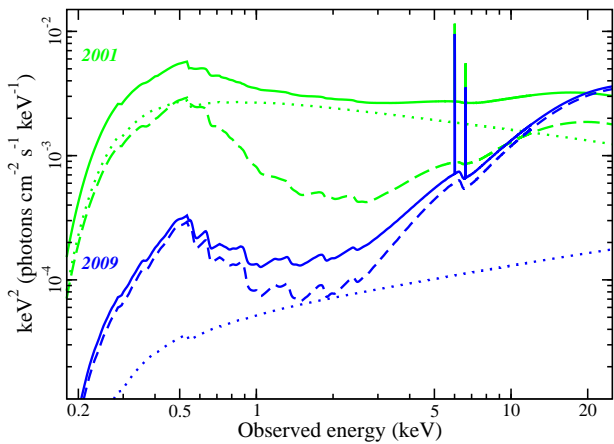}}
\scalebox{0.32}{\includegraphics{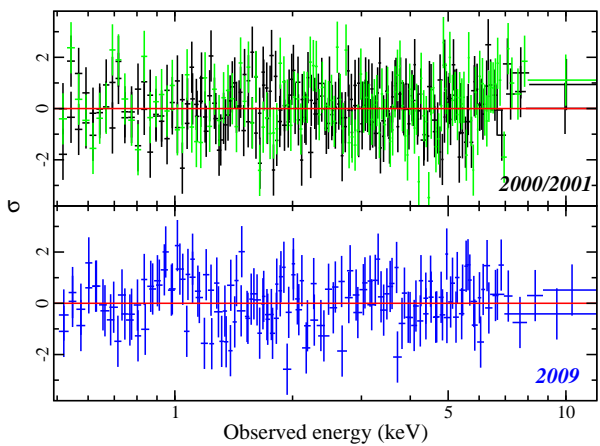}}}
\caption{Upper panel: The ionised reflection model used to fit the high-state (2001) and 
low-state (2009) spectra of \pg08.  The 2000 model is very similar to the 2001 and is not
shown for clarity.  The dotted curve and dashed curve are the power law and reflection components,
respectively.  The solid curve is the combined model.  The reflection component dominates the
X-ray band in the low-state.
Lower panel: Spectral residuals (in sigma) based on this model (see text and Table~\ref{tab:fits} for details)
for the high-state (top) and low-state (bottom).
}
\label{fig:reffit}
\end{figure}
% --------------------------------------------------------------------------

The model provides a good fit to the spectra at all three epochs ($\chidof=1.05/474$; Figure~\ref{fig:reffit}),
and can be considered self-consistent.  While there are several parameters that fluctuate
from epoch-to-epoch the primary driver is the diminishing of the power law normalisation from
high- to low-flux.   
Indeed, during the X-ray weak state the X-ray spectrum of \pg08\ is dominated by the reflection
component (Figure~\ref{fig:reffit}, upper panel).   We estimate the reflection fraction ($R$) at
each epoch by comparing the expected model above $10\keV$ (where ionisation and absorption effects
are minimum) with the {\tt pexrav} model in {\tt XSPEC} (Magdziarz \& Zdziarski 1995).  During the
high-flux state $R\approx0.8$, commensurate with an isotropic emitter.  In the X-ray weak state,
$R>>1$ indicating that extreme effects, such as light bending, may be at play.

The models also show a flattening of the power law slope from high- to low-state.
While some variability is expected in the shape of the power law with time, the change measured
here is rather significant and difficult to reconcile.
Similar behaviour was noted in another AGN, Mrk~79 (Gallo \et 2010).

% --------------------------------------------------------------------------
\begin{figure}
\rotatebox{270}
{\scalebox{0.32}{\includegraphics{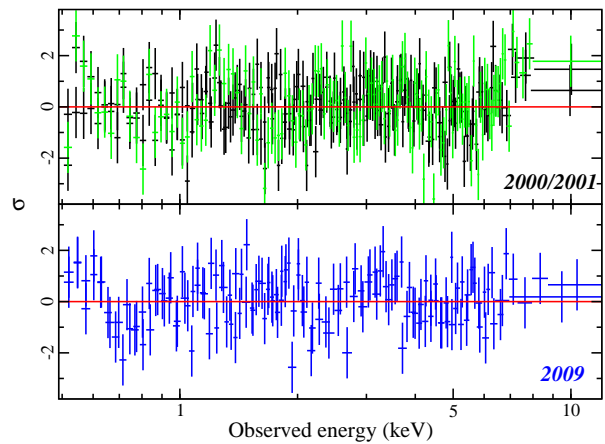}}
\scalebox{0.32}{\includegraphics{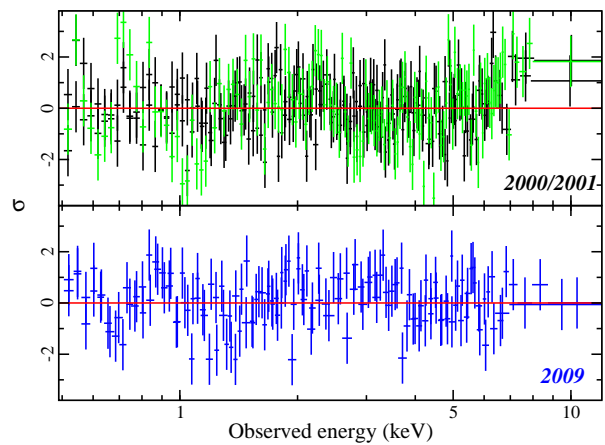}}}
\caption{
Upper panel: Spectral residuals (in sigma) from modeling the high-state (top) and low-state (bottom)
of \pg08\ with a neutral double partial covering (see text and Table~\ref{tab:fits} for details). 
Lower panel: Same as upper panel, but for the ionised double partial covering model.
}
\label{fig:pcfit}
\end{figure}
% --------------------------------------------------------------------------

\subsection{Neutral partial covering }
\label{sect:npc}
We consider a scenario where the
intrinsic power law continuum is partially obscured by a neutral absorber along the 
line-of-sight (e.g. Grupe \et 2004, Gallo \et 2004).  
In this picture, the observed spectrum is the combination of direct emission from the power law
emitter and a highly obscured component.  Since a significant amount of time passes between observations,
we allow for both intrinsic continuum changes (i.e. changes in the power law shape and luminosity) and
changes in the
characteristics of the absorber [i.e. column density ($\nh$) and covering fraction ($C_f$)]. 
The simplest case of a single absorber results in a poor fit ($\chidof=2.09/479$) and
supports the addition of a second partial covering absorber (i.e. neutral double
partial covering).
The double partial covering model is akin
to two distinct absorbing regions or a column density gradient along
the line-of-sight (e.g. Tanaka \et 2004).  

The fit is significantly improved ($\chidof=1.16/473$) with the addition
of the second absorber.  The model parameters and fit quality are shown in Table~\ref{tab:fits}
and Figure~\ref{fig:pcfit} (top panel), respectively.
The two absorbers have similar characteristics during the high-state observations (2000 and 2001) and together
obscure about $70$~per cent of the power law light.  To account for the low-flux state the column density and
covering fraction of both absorbers increase, and together diminish the source flux 
in the $0.5-10\keV$ band
by about $93$~per cent.
We also note that the intrinsic power law flux dropped by about $67$~per cent in 2009 from its 2001 high.
Noteworthy is the particularly steep spectrum that is required for the continuum ($\Gamma\approx3$) 
at each epoch.

\begin{figure}
\rotatebox{270}
{\scalebox{0.32}{\includegraphics{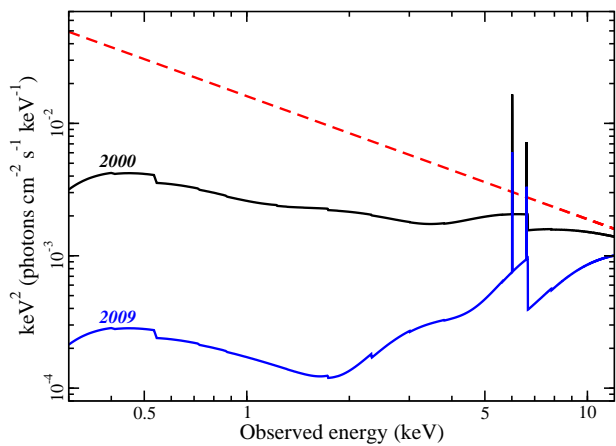}}}
\caption{
The neutral partial covering model assuming a constant primary power law (red dashed line) at all epochs.
The changes in the observed spectrum are due to variations in the absorber covering fraction and
column density.  For clarity, only the 2000 and 2009 models are shown.
}
\label{fig:pclum}
\end{figure}

A statistically worse, but not unreasonable fit, can be obtain by maintaining the shape and luminosity
of the primary power law constant at all three epochs ($\chidof=1.26/477$; Figure~\ref{fig:pclum}).
In this case, variations in the nature (covering fraction and column density) of the absorber
account for the long-term variations in the shape and flux of the spectrum.  The intrinsic
luminosity of the power law in the $0.5-10\keV$ band is $5.2\times10^{44}\ergps$.  The power law
photon index remains steep ($\Gamma=2.93\pm0.03$).
% --------------------------------------------------------------------------
% --------------------------------------------------------------------------

\subsection{Ionised partial covering }
\label{sect:ipc}

A modification of the partial covering model discussed above is to consider an absorber that
is ionised (Reeves \et 2008) rather than neutral.  
The ionised material will preferentially absorb intermediate energy X-rays
giving rise to the different spectral slopes on either side of $E\approx2\keV$.  

A single ionised absorber partially covering the primary continuum is a considerably better fit than
the single neutral absorber ($\chidof=1.36/476$), but significantly worse than the neutral double
partial covering and ionised reflection models.  
The addition of a second ionised absorber was attempted to improve the fit.  
In initial fits the ionisation parameter of the second absorber was
comparable at all three epochs so it was linked in subsequent fits.
The improvement over the single ionised absorber was substantial ($\delchi=160$ for 7 additional parameters).  
The fit residuals are displayed in Figure~\ref{fig:pcfit}.
Notably, the ionisation of the absorber does not follow the predicted flux of the continuum
as would be expected if the power law is the sole source of ionisation for the absorber
and if the absorber was of high density.

The two absorbers collectively obscure about $50$~per cent of the power law continuum in the high-state and
about $90$~per cent in the low-flux state.  
The moderately low ionisation parameters predicted with this model will produce a spectrum rich in
absorption features that should be detectable in high-resolution spectra even in the high-state.
An analysis of the 2000 \xmm\ data (including the RGS) by Pounds \et (2003) resulted in claims of highly blue-shifted
absorption lines due to a high-velocity outflow, but was later dismissed by Brinkmann \et (2006).

Similar to the neutral partial covering, the long-term variations could be modelled with a
constant power law and varying absorption parameters.  The fit is slightly worse than that presented
above ($\chidof=1.16/473$) and the intrinsic power law is again steep ($\Gamma=2.77\pm0.30$).
 
\section{Discussion } 
\label{sect:dis}

In the X-ray weak state \pg08\ shows considerable spectral hardening and
curvature compared to the relatively smooth spectrum during the bright state.  
Blurred ionised reflection models fit the multi-epoch spectra very well in a self-consistent manner. 
Partial covering models, with either neutral or ionised absorbers, also work reasonably well.
The reflection and absorption models appear spectroscopically similar in the $0.5-10\keV$ band,
but each model appears different at higher energies and predict different temporal behaviour.  

In the blurred reflection model the principle difference between the high- and low-state is the 
prominence of the direct power law continuum reaching the observer.  There is more than a factor
of $10$ difference in the $0.5-10\keV$ power law flux between the high- and low-state.  In fact, during the
low-state the reflection component dominates the EPIC spectrum and the measured reflection fraction is
$R>>1$.  The attenuation of the direct power law flux could be attributed to obscuring of the component
(e.g. Fabian \et 2002) or by light bending effects (Miniutti \& Fabian 2004).

In addition, the reflection models also indicate variations in the flux and ionisation of the
reflection component.  These changes are manifested in a model-independent way in the difference 
spectrum (Figure~\ref{fig:diffs}).  Rather than being a simple power law (characteristic of normalisation
changes alone), the difference spectrum has a soft excess.  Ionisations changes in the reflection
component would modify the soft excess.

\begin{figure*}
\begin{center}
\begin{minipage}{0.48\linewidth}
\scalebox{0.32}{\includegraphics[angle=270]{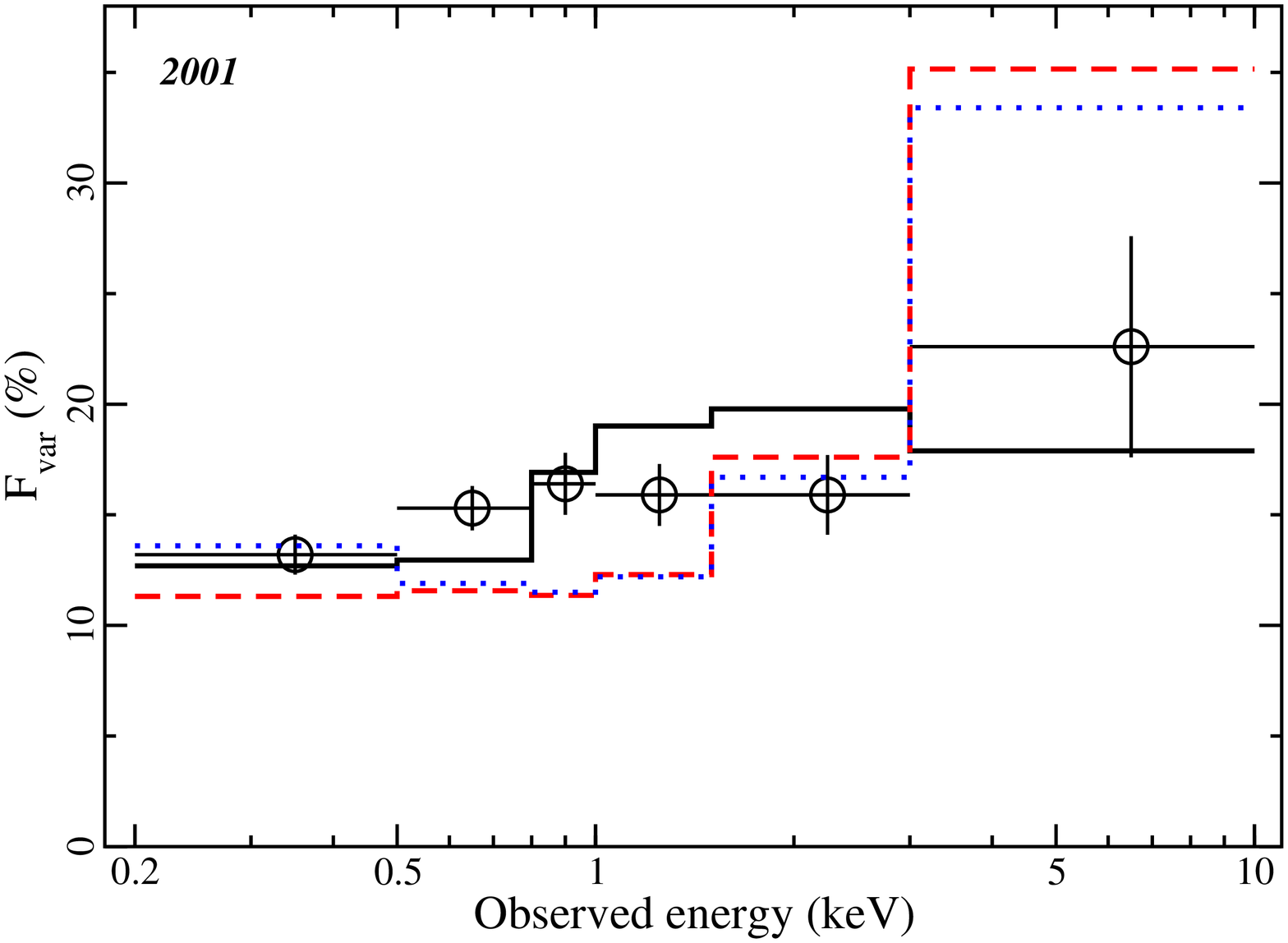}}
\end{minipage}  \hfill
\begin{minipage}{0.48\linewidth}
\scalebox{0.32}{\includegraphics[angle=270]{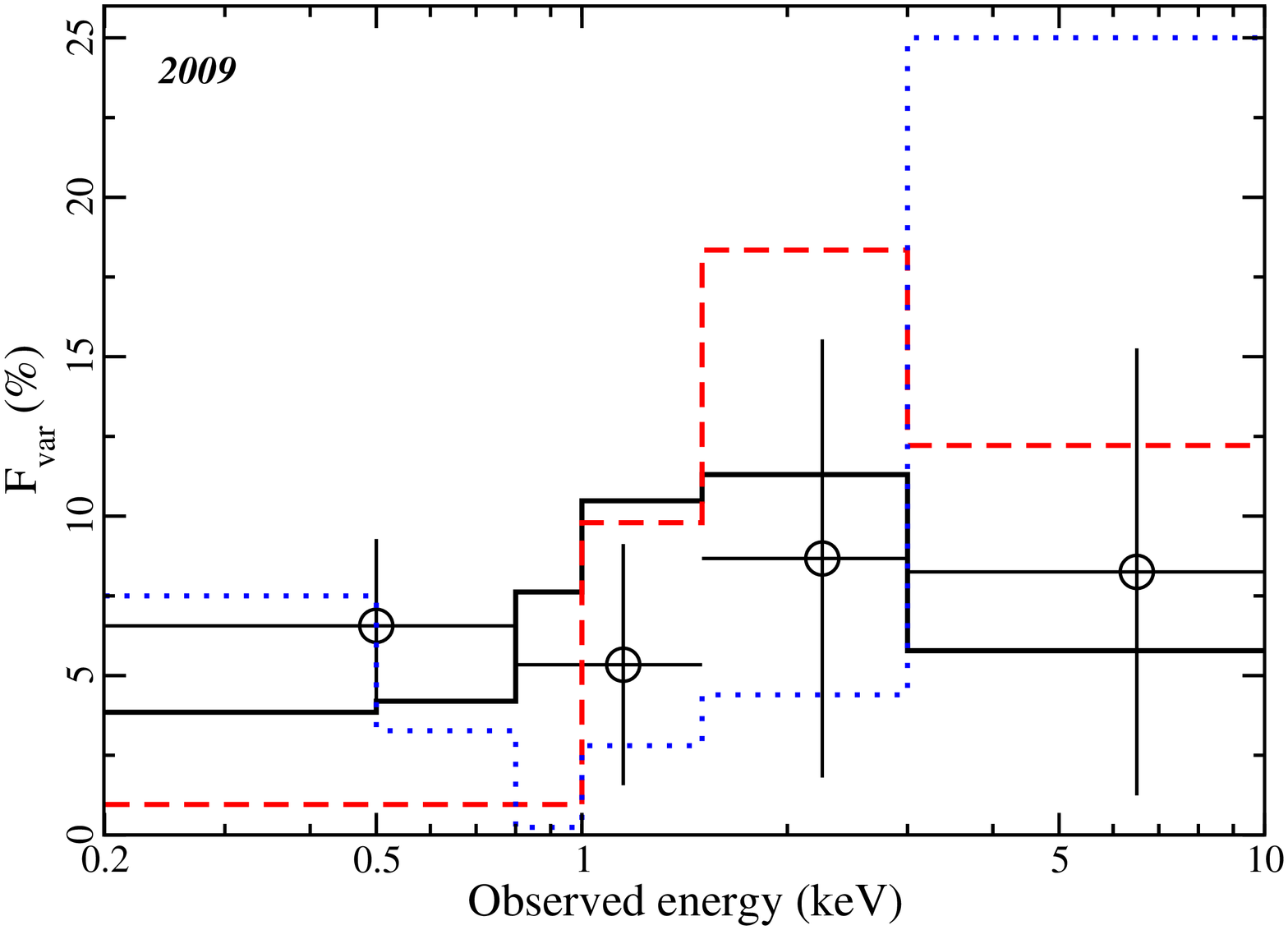}}
\end{minipage}
\end{center}
\caption{The normalised rms spectrum calculated for the 2001 (left panel) and
2009 observations (right panel).  On the assumption that only the power law normalisation
fluctuates, the expected $\fvar$ based on the blurred reflection
(black solid), double neutral partial covering (red dashed) and double ionised partial
covering (blue dotted) models are shown.  
The single ionised absorber produces a similar $\fvar$ as the double ionised absorber
and is not shown.
The light curves used are in $1000\s$ bins.
}
\label{fig:fvar}
\end{figure*}

The light bending
scenario makes specific predictions about the rapid temporal behaviour of the power law and reflection components in
various flux states that can in turn be tested. 
In examining the 2001 $\fvar$ spectrum (Figure~\ref{fig:fvar} left panel) we consider a situation where
the power law component fluctuates in brightness and dominates the variability (i.e. the reflection component
varies very little in comparison).  Such conditions are expected from the light bending model when the AGN
is in the bright state (Miniutti \& Fabian 2004).
Such a model agrees well with the data of \pg08\ (Figure~\ref{fig:fvar} left panel).
% --------------------------------------------------------------------------
% --------------------------------------------------------------------------

Likewise, the expected $\fvar$ based on the partial covering models is also shown in
Figure~\ref{fig:fvar}.  The models are adequate given the signal-to-noise, but both models clearly
miss the data in some energy bins.  The absorption models likely require variability of an additional
parameter to reconcile the $\fvar$.  Gierli\'nski \& Done (2006) consider changes in the
ionisation parameter of the absorber (in addition to power law flux) to reconstruct
the $\fvar$ of some sources that show the recognised peak at intermediate energies.

In a similar fashion we conducted the same exercise for \pg08\ in the low-flux state
(right panel Figure~\ref{fig:fvar}).  The spectrum is of lower quality than the 2001 observation as
the amplitude of the variations is low and the observation is short,
and consequently no model can be ruled out with statistical certainty.  However, we note that the models
predict significantly different behaviour and with a longer low-flux observation (and 
higher signal-to-noise) the models could potentially be distinguished.

If due to partial covering, the
simultaneous UV/X-ray data could potentially place some restrictions on the geometry of the system
depending on the nature of the absorber.
The UV observations (all longward of the Lyman limit) do not appear to exhibit diminishing flux as the 
X-rays do. 
The fact that we do not see UV variability simultaneous with the X-rays implies that any
partial covering absorber, if present, has to be dust-free and
confined to the X-ray emitting region.  
This requires the existence of very dense, cold blobs that are a substantial
fraction of the source size, occupying a small region close to the source.
Ostensibly this would imply the presence of a strong iron fluorescence line that is not
observed in the data.  Such a strong feature would be detectable unless the covering fraction was
small or in a rather contrived geometry (see Reynolds \et 2009 and Miller \et 2010 for opposing opinions).

The UV spectrum of \pg08\ also displays relatively weak C{\textsc iv} absorption (Brandt \et 2000).
The feature is expected to be much stronger if the X-ray weak state of \pg08\ were due to absorption
from a BAL-related phenomenon (Brandt \et 2000).  UV spectroscopy of \pg08\ when it is in a X-ray weak
state would confirm the absence of a BAL absorber.

The inferred intrinsic photon index measured in \pg08\ is steeper than
the canonical value of $\Gamma\approx1.9$ that is normally adopted (e.g. Nandra \& Pounds 1994).  
This seems to
be independent of the model used and the flux-state of the AGN (except for the 2009 blurred reflection
interpretation, see Sect.~\ref{sect:refl}).  There are potential explanations for this, for example
if due to Comptonisation \pg08\ could have a cooler corona than the average AGN leading to a
steeper spectrum.  It is difficult to put this in to context of the NLS1/BLS1 nature of \pg08.
NLS1s are known to exhibit steeper spectra than BLS1 (e.g. Boller \et 1996; Brandt \et 1997),
but this is only based on the observed spectrum that is modified by reflection and warm absorption.
A true comparison of the intrinsic NLS1 and BLS1 photon index to examine for physical difference
between the two classes is still lacking.  However, in the near future, as BAT detections become
more significant and hard X-ray imaging becomes possible, we will be in position to answer this question.

Each model has unique spectral signatures that could be discerned with higher spectral 
resolution or broader energy bandpass.  The ionised absorber predicts absorption features in the
spectra at $6-8\keV$ due to various transitions in iron.  
Such features may be lost in the EPIC data due to the moderate spectral 
resolution.  However, Shu \et (2010) present the \chandra\ HETG spectrum of \pg08\
when it was in a comparable flux state as the 2000 and 2001 \xmm\ observations.  
The HETG spectrum 
is of modest quality, but shows no
indication of absorption features in the iron region (see figure 1 of Shu \et).
A higher signal-to-noise HETG spectrum could potential constrain such features.

Finally, all three models make vastly different predictions of the spectral shape and flux above
$10\keV$.  The absorption models have steeper intrinsic spectra ($\Gamma = 2.4-3$) and a weaker
reflection component, whereas the blurred reflection model predicts a much flatter spectrum.  
The reflection model also predicts a $10-20\keV$ flux that is twice as large
in the high-state and $4-5\times$ greater in the low-state than the absorption models.
Based on our models, the brightest flux reached by \pg08\ in the $14-195\keV$ band is
$\sim10^{-11}\ergpscmps$ and fluctuates to lower fluxes.  This peak value is still a factor of 2 
lower than the current detection limit of the \swift\ BAT
survey (Tueller \et 2010).  

Through multi-epoch X-ray observations the reflection-dominated low-state spectrum of \pg08\ 
appears more clearly.  Similar observations of several other AGN also support the notion that
the AGN X-ray low-flux state may be due to a diminished power law that in turn reveals the underlying reflection
spectrum.
Multi-epoch observations are necessary to distinguish between competing models.  Placing constraints on the
nature of the long-term spectral variability is educative and feasible with current instruments.

\section{Summary } 

A $15\ks$ \xmm\ ToO observation of the Seyfert~1 \pg08\ was triggered when it was discovered
with \swift\ in 2009 that the AGN was in an X-ray weak state.
Here we conducted an analysis of the X-ray spectrum and the long-term variability by
combining data of the AGN from previous epochs when it was in a bright state.

\pg08\ is fitted with various physical models (blurred reflection and partial covering).  While no
model is conclusively dismissed the blurred reflection model nicely describes the long- and short-term
variability in a consistent manner.  The two classes of models predict distinguishing characteristics
that could be revealed with deeper observations (i.e. high
signal-to-noise and broader energy band) of \pg08\ in the low-flux state with current X-ray observatories.
In the future, such observations will be much simpler.  
The high-energy imaging capabilities and calorimeter resolution of {\it Astro-H} and {\it IXO} will provide 
the opportunity to
simultaneously measure the broadband spectrum while achieving high spectral resolution below $10\keV$,
and {\it NuSTAR} will be the first to image the high-energy band.

% --------------------------------------------------------------------------
% --------------------------------------------------------------------------

\section*{Acknowledgments}

Thanks to Neil Gehrels for approving the \swift\ ToO request.
\swift\ at PSU is supported by NASA contract NAS5-00136.
The \xmm\ project is an ESA Science Mission with instruments
and contributions directly funded by ESA Member States and the
USA (NASA).  We are grateful to the \xmm\ observing team for preparing and
activating the ToO.  We thank the referee for helpful comments and input.
This research has made use of the NASA/IPAC Extragalactic
Database (NED) which is operated by the Jet Propulsion Laboratory,
Caltech, under contract with the National Aeronautics and Space
Administration.
This research was also supported by NASA contracts NNX07AH67G and NNX09AN12G
(D.G.).

%\pagebreak

%\appendix
%\section{Analysis of EPIC calibration data}
%\label{app:cal}

%\clearpage

\bsp
\label{lastpage}
\end{document}